\documentclass[preprint,showpacs,amsmath,amssymb,aps,prc,superscriptaddress]{revtex4}
\usepackage{epsfig}
\begin{document}
\preprint{APS/123-QED}

\title{INFLUENCE OF NEUTRON ENRICHMENT ON DISINTEGRATION MODES OF COMPOUND 
NUCLEI}

\author{E. Bonnet$^{ 1}$, J.P. Wieleczko$^{ 1}$, A. Chbihi$^{1}$, J.D. 
Frankland$^{1}$, J. Moisan$^{1}$, F. Rejmund}

\affiliation{GANIL, CEA et IN2P3-CNRS, B.P. 55027, F-14076, Caen Cedex, France}
 
\author{J. Gomez del Campo$^{ 2}$, A. Galindo-Uribarri$^{2}$, D. Shapira}

\affiliation{Physics Division, Oak Ridge National Laboratory, Oak Ridge, TN 37831, 
USA}

\author{M. La Commara$^{3}$, B. Martin$^{3}$. D. Pierroutsakou$^{3}$,
M. Romoli$^{3}$, E. Rosato$^{3}$, G. Spadaccini$^{3}$. M. Vigilante}

\affiliation{Dipartimento di Scienze Fisiche and INFN, Universit\`a di Napoli 
"Federico II", I80126, Napoli, Italy}

\author{ S. Barlini$^{4}$, R. Bougault $^{4}$, N. Le Neindre$^{4}$, M. 
Parlog$^{4}$, B. Tamain}

\affiliation{LPC, IN2P3-CNRS, ENSICAEN et Universit\'e, F-14050, Caen Cedex, 
France}

\author{ C. Beck}

\affiliation{Institut Pluridisciplinaire Hubert Curien - D\'epartement de
Recherches Subatomiques, UMR7178, 
IN2P3-CNRS et Universit\'{e} Louis Pasteur (Strasbourg I), 23 rue du Loess, 
B.P. 28, F-67037 Strasbourg Cedex 2, France }

\author{ B. Borderie $^{6}$, M.F. Rivet}

\affiliation{IPNO, IN2P3-CNRS, F-91406, Orsay Cedex, France}

\author{ R. Dayras$^{7}$, L. Nalpas}

\affiliation{CEA, IRFU, SPhN, CEA/Saclay, F-91191, Gif sur Yvettes Cedex, 
France}

\author{ G. De Angelis$^{8}$, T. Glodariou$^{8}$, V. Kravchuk}

\affiliation{INFN, LNL, I 35020 Legnaro (Padova) Italy}
  
\author{Ph. Lautesse}

\affiliation{IPNL, IN2P3-CNRS et Universit\'e, F-69622, Villeurbanne Cedex, 
France}

\author{ A. D. Onofrio}

\affiliation{Dipartimento di Scienze Ambiantali, Seconda Universit\`a di 
Napoli, I 81100, Caserta, Italy}

\author{R. Roy}

\affiliation{Laboratoire de Physique Nucl\'eaire, Universit\'e de Laval, 
Qu\'ebec, Canada}


\date{\today}

\begin{abstract} 

Cross sections, kinetic energy and angular distributions of fragments with charge 6$\le$Z$\le$28 emitted in $^{78,82}$Kr+$^{40}$C at 5.5 MeV/A reactions were measured at the GANIL facility using the INDRA apparatus. This experiment aims  to investigate the influence of the neutron enrichment  on the decay mechanism of excited nuclei. Data are discussed in comparison with predictions of transition state and Hauser-Feshbach models.

\end{abstract}
\maketitle

\newpage

\section{Introduction}

Nuclei under extreme conditions (temperature, spin, neutron to proton ratio N/Z, density) have been the focus of experimental and theoretical investigations. The fusion process between heavy ions colliding at incident energies around the Coulomb barrier is well suited to study hot and rotating compound nuclei. These excited nuclei decay through a variety of channels such as neutrons, light charged particles, fission and production of intermediate mass fragments (IMF's) with charge Z$\ge$3. The neutron enrichment of the compound nuclei is expected to influence the competition between these decay modes. In this contribution we present new data on the production of fragments with 6$\le$Z$\le$28 formed in $^{78,82}$Kr+$^{40}$Ca fusion reactions at 5.5 MeV/A incident energy. 

\newpage

\section{Experimental results.}
The experiments were performed at the GANIL facility and $^{78,82}$Kr beams were used to bombard a self-supported $^{40}$Ca target of 1mg/cm2 in thickness. Fragments were measured with the 4$\pi$-INDRA array \cite{1}. For 3$\le\theta_{lab}\le$ 45, each module comprises an ensemble made of ionization chamber, silicon detector and CsI. For fragments with 6$\le$Z$\le$28, this angular range covers the forward hemisphere in the centre of mass frame. The energy calibration was deduced from the Kr+Ca elastic scattering and energy loss tables.
For both systems, the centre of mass average velocity of fragments is roughly constant as a function of the emission angle and agrees rather well with the Viola systematic\cite{2}. The angular distribution d$\sigma$/d$\theta_{cm}$ of fragments is almost isotropic in the measured angular range and the sum of the charges of the two biggest fragments is close to the charge of the compound nucleus. All these features are compatible with binary emission from a compound nucleus. The absolute cross-sections were obtained from the normalization to the elastic scattering measured at a laboratory angle for which the scattering is given by the Rutherford formula. 
The measured charge distributions are shown in fig.1a for $^{82}$Kr+$^{40}$Ca. The cross-section distributions present a shape with a maximum around half of the charge of the compound nuclei. This indicates that elements with Z$\ge$14 come mainly from a symmetric fission process induced by  high transferred spins. 
Besides this feature, for light fragment (Z$\le$10) a strong even-odd staggering is visible, and this effect is still present for higher Z with a smaller amplitude. Similar features are observed for the $^{78}$Kr+$^{40}$Ca reaction. However the yields for symmetric splitting are about 30\% higher than for the $^{82}$Kr+$^{40}$Ca system. Cross section for odd-Z fragments is higher for the neutron rich nuclei system and for even-Z fragments it is higher for the neutron poor nuclei system. Fig. 1b shows the charge dependence of the ratio R= $\sigma^{118Ba}_{Z}/\sigma^{122Ba}_{Z}$. R decreases roughly from 1.25 for Z=6 down to 0.75 for Z=7. For odd Z, the ratio increases up to Z=21 and reaches a kind of plateau. For even Z, R decreases from Z=6 to Z=10 and then increases to reach the same kind of plateau as for odd-Z. The excitation energy and the maximum angular momentum stored in the compound nucleus are expected to be very similar in both reactions. Thus, the observed effects are probably linked to the difference in the neutron enrichment of the compound nucleus. It is worth noticing that the Z=6 cross section is comparable to the one infered in \cite{3}

\newpage

\section{Comparisons with statistical model calculations.}

Statistical decay calculations were performed using the Monte Carlo code 
GEMINI \cite{4}. All channels are considered within the Hauser-Feshbach (for Z$\le$2) and transition state (for Z$\ge$3 ) approaches. The key ingredient is the conditional saddle configuration (or conditional barrier) which is constrained by the mass asymmetry. The saddle conditional energy for different mass (or charge) asymmetry was deduced from Sierk's model \cite{5}. In the present work, the fusion evaporation cross section is not yet available but, for this kind of studied reaction, we know that the fission cross section is sensitive to the highest angular momentum leading to fusion mechanism. Thus, the maximum angular momentum $J_{max}$ was considered as a free parameter to reproduce the yields around the symmetric splitting. All the calculations presented here have been performed using the standard parameters of the GEMINI code and a level density parameter a=A/8. 
Results are reported in fig.1a for the $^{82}$Kr+$^{40}$Ca and fig. 1b for the ratio $\sigma^{118Ba}_{Z}/\sigma^{122Ba}_{Z}$.  In both figures, the line represents the best fit obtained with the indicated $J_{max}$. The shape of the charge distribution around symmetric fission is relatively well reproduced, while the model predicts a smooth behaviour for Z$\le$14 at variance with the data. For fragments with 6$\le$Z$\le$12, the calculation overestimates the measured cross-sections. Various attempts at changing the $J_{max}$ value lead to the conclusion that the present version of the code fails to explain the relative population of the IMF's and symmetric fission.
The key ingredient governing the competition between channels is the barrier. Besides the even-odd effect, the observed disagreement would indicate a failure of the Sierk  barriers and/or the transition state picture for large asymmetry and large angular momentum. Indeed, in the transition state picture the asymmetry at saddle configuration is frozen all along the path to scission. For medium-mass compound nuclei at high spin as those studied in this work, data presented in fig.1a suggest that saddle and scission configurations are no longer degenerate. Then, the surface of the potential energy for high angular momentum would be narrower, and the charge distribution too, leading to an increase of the relative yields between the IMF's and symmetric fission.
Coming back to the even-odd effects in cross-sections, the GEMINI code allows to include the washing out of the shell and pairing effects with increasing temperature.  Standard calculations, as such  those shown in fig.1, correspond to exclude any structure effect for Z$\ge$3 . We made an attempt assuming a phenomenological description of pairing. In that case, the even-odd effect  predicted by the model has a smaller amplitude with respect to the data. Moreover, the ratio $\sigma^{118Ba}_{Z}/\sigma^{122Ba}_{Z}$ reproduces experimental data for light fragments with odd Z and strongly underestimates the yields of even Z. The agreement becomes satisfactory for Z$\ge$14 . It is worth noticing that for Z$\ge$14 the shape of the Z distribution is well reproduced. From these comparisons, we can conclude that the way the pairing is included requires more investigations. As noticed above, the calculated yields of the IMF's are overestimated, possibly due to an underestimation of the barrier. Thus, in the model, the available excitation energy available after the splitting would be higher, leading to higher excitation energy in the IMF's. The main consequence would be a decreasing of the even-odd effect due to an overestimation of the calculated temperature of the IMF's. To explore this scenario we performed some calculations using of a different approach of the statistical model. 
\begin{figure}[th]
\centerline{\psfig{file=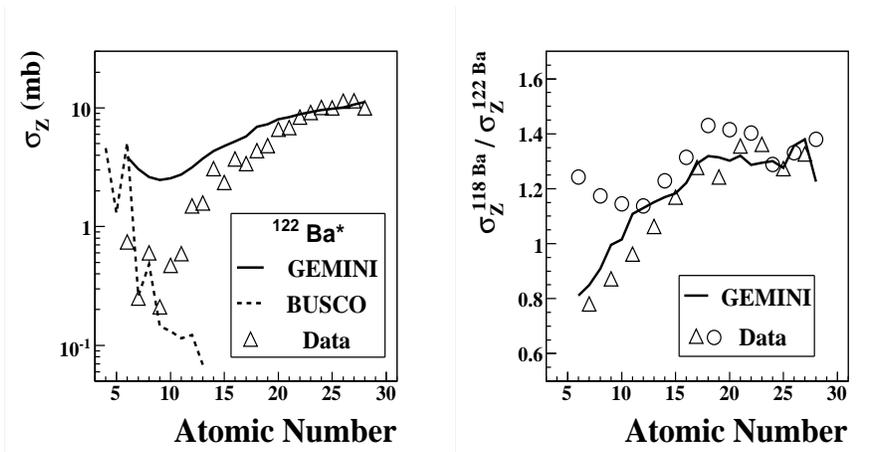,width=12cm,height=6cm}}
\vspace*{8pt}
\caption{(a) Experimental cross sections for fragments emitted in $^{82}$Kr+$^{40}$Ca (triangles), compared to GEMINI ($J_{max}$=66$\hbar$, line) and BUSCO ($J_{max}$=45$\hbar$, dashed line). (b) Experimental ratio $\sigma^{118Ba}_{Z}/\sigma^{122Ba}_{Z}$ (triangles and circles stand respectively for odd-Z and even-Z fragments) compared to GEMINI calculations (line).}
\end{figure}
The measured even-odd effect in IMF  yields suggests that they are emitted relatively cold or close to the threshold of particle emission otherwise the picture would have been blurred by secondary emission. In the BUSCO model \cite{6} the decay of compound nuclei includes all open channels from Z=0 to Z=20. Partial widths are calculated within the Hauser-Feshbach framework. Ground and excited states of each emitted fragment were included. An attempt was made assuming $J_{max}$=45$\hbar$. From the comparison to the data the model fails to reproduce the Z distribution, the calculated cross section decreasing strongly as Z increases at variance with data. Fixing a value for $J_{max}$ in order to reproduce cross section for Carbon leads to underestimate the yields of Z=10 by about three orders of magnitude. The calculated ratio $\sigma^{118Ba}_{Z}/\sigma^{122Ba}_{Z}$ is completely wrong and is not shown in fig.1b.
Further experimental investigations will be performed to analyse the light charged particles emitted in coincidence with IMF's. This will inform on the excitation energy stored in the IMF's and on the level density parameter. Measurement of the evaporation residue cross section will allows to deduce the maximum angular momentum for fusion process. To improve the modelization, refinements are needed to describe the disintegration process in an excitation range where structure effects influence strongly the phase space available for a statistical description.

\end{document}